\documentclass[conference]{IEEEtran}
\IEEEoverridecommandlockouts

\usepackage{cite}
\usepackage{amsmath,amssymb,amsfonts}
\usepackage{algorithmic}
\usepackage{graphicx}
\usepackage{textcomp}
\usepackage{diagbox}
\usepackage{makecell}
\usepackage[table]{xcolor}
\def\BibTeX{{\rm B\kern-.05em{\sc i\kern-.025em b}\kern-.08em
    T\kern-.1667em\lower.7ex\hbox{E}\kern-.125emX}}

\begin{document}

\title{Asymmetric Transistor Aging Under Thermal Gradients}

\author{\IEEEauthorblockN{1\textsuperscript{st} Alex Grinshpun}
\IEEEauthorblockA{\textit{The Institute of Electrical Engineering and Applied Physics} \\
\textit{The Hebrew University, Jerusalem, Israel}\\
alex.grinshpun@mail.huji.ac.il}
\and
\IEEEauthorblockN{2\textsuperscript{nd} Firas Ramadan}
\IEEEauthorblockA{\textit{The Faculty of Electrical and Computer Engineering} \\
\textit{Technion, Haifa, Israel}\\
firasramadan@campus.technion.ac.il}
\and
\IEEEauthorblockN{3\textsuperscript{rd} Freddy Gabbay}
\IEEEauthorblockA{\textit{The Institute of Electrical Engineering and Applied Physics} \\
\textit{The Hebrew University, Jerusalem, Israel}\\
freddy.gabbay@mail.huji.ac.il}
}

\maketitle
\begin{abstract}
Advanced technology nodes improve performance but increase susceptibility to BTI-induced aging. Using aging-aware circuit simulations, we show that realistic on-chip thermal gradients can cause significant asymmetric transistor degradation in 16 nm FinFET circuits, leading to setup and hold time violations. These results highlight the necessity of incorporating thermal-gradient-aware aging analysis into timing signoff to ensure chip reliability.
\end{abstract}


\section{Introduction}
As the semiconductor industry continues to scale devices for higher performance and density, ensuring long-term reliability in advanced nanoscale technologies has become increasingly challenging ~\cite{liu2020new,aging-5nm}. Modern chips fabricated in advanced technology nodes are highly susceptible to transistor aging, a degradation mechanism that increases the threshold voltage (Vth), reduces the saturation current (Isat), and consequently slows down circuit timing. This degradation is mainly driven by mechanisms such as Hot Carrier Injection (HCI) and, most critically, Bias Temperature Instability (BTI). BTI causes a gradual increase in the transistor threshold voltage, leading to reduced drive current, slower switching, and eventual timing failures \cite{ref11,SA-Paper,ref38,ref39}, which might lead to system failures and silent data corruption (SDCs) \cite{agarwal2007circuit,anik2020effect,amrouch2015techniques,amrouch2015connecting,dixit2021silent,wang2023understanding,dong2021evolution,serebryany2021silifuzz,bacon2022detection,dixit2022detecting}. 

Asymmetric aging refers to the non-uniform degradation of transistors within an integrated circuit (IC). This phenomenon arises from several factors, most notably spatial variations in switching activity and bias duty cycle across the chip \cite{ref11,SA-Paper,ref37}. Power-saving mechanisms such as clock gating and power gating can further exacerbate this imbalance by subjecting large circuit regions to prolonged static voltage stress \cite{ref38,ref39}. As a result, delay degradation becomes uneven, with rarely toggled or always-on paths aging significantly faster than their high-activity counterparts.

Beyond workload effects, a critical but underexplored source of asymmetric aging is the presence of on-chip thermal gradients. Modern SoCs exhibit spatial temperature gradients reaching 10–15 °C \cite{Gradient}. Because BTI accelerates exponentially with temperature \cite{ref1_new_2}, even small temperature gradients can substantially affect aging rate. Thus, transistors placed in different thermal regions can age at different rates, compounding the asymmetry beyond workload-induced effects.

Despite this, most reliability and EDA timing flows still assume a uniform worst-case temperature, overlooking how spatial thermal variation shapes transistor aging degradation. In this paper, we present a study of asymmetric transistor aging under realistic thermal gradients in a 16 nm process node. We have conducted aging simulations using TMI-based aging libraries and Cadence Spectre circuit simulator \cite{spectre}. Our simulations show that temperature gradients of 10–15 °C can induce significant non-uniform BTI aging, leading to both setup and hold violations that conventional signoff methods fail to capture. Our results underscore the importance of incorporating thermal-gradient effects into standard signoff to ensure reliable operation. The remainder of this paper is organized as follows: Section \ref{sec:two} reviews related work, Section \ref{sec:three} presents our analysis, and Section \ref{sec:four}  concludes the paper.
 

\vspace{-5pt}
\section{Related Studies} \label{sec:two}
\cite{ref37} investigates asymmetric aging in microprocessors and proposes microarchitectural techniques to mitigate uneven degradation. \cite{SA-Paper} examines BTI-induced asymmetric aging in systolic arrays and its impact on DNN accuracy, introducing hardware-level countermeasures. \cite{ref38} studies aging effects on clock-tree structures and presents design-flow extensions to reduce timing violations. Recent GPGPU studies show that BTI-induced degradation affects execution units and propose mitigation techniques for asymmetric aging \cite{Micdat2023}. Work on unmanned vehicles (UxVs) introduces predictive-maintenance frameworks to detect aging-induced faults in safety-critical systems \cite{PM} .\cite{MissionProfile} demonstrates that realistic workload and mission profiles strongly influence aging behavior and that workload-aware simulations can reduce overdesign.


\section{Simulation Analysis of Asymmetric Aging Induced by Thermal gradients} \label{sec:three}
\begin{figure}[h]
    \centering
    \includegraphics[width=0.7\linewidth, height=100pt]{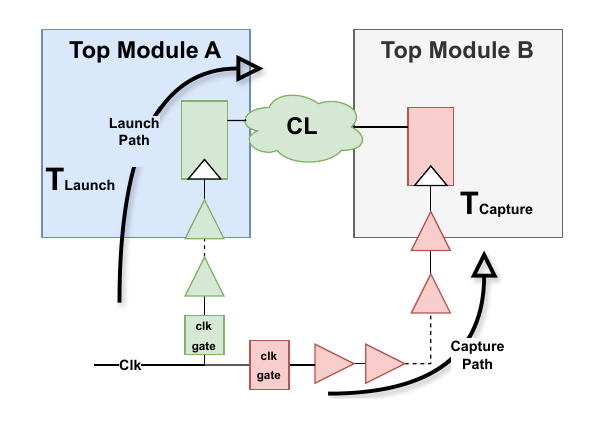}

\vspace{-10pt}
    \caption{Top-Level Modules - Scenario A}
    \label{fig:top_module2}
\end{figure}

\begin{figure}[h]
    \centering
    \includegraphics[width=0.7\linewidth, height=100pt]{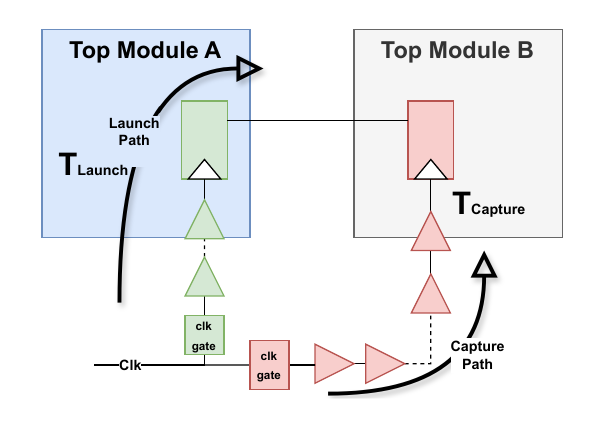}
        \vspace{-10pt}

    \caption{Top-Level Modules - Scenario B}
    \label{fig:top_module}
\end{figure}

Transistor aging due to BTI is strongly temperature-dependent, with higher junction temperatures accelerating threshold-voltage shifts and performance degradation\cite{ref1_new_2}. Even moderate on-die thermal differences can lead to divergent aging rates; temperature gradients of 10–15$^\circ$C are common in modern SoCs and are sufficient to induce asymmetric degradation across nearby units\cite{Gradient}. As technology nodes scale and power density rises, this sensitivity becomes even more pronounced. Such thermal gradients arise from multiple factors. Modules with high logic density inherently dissipate more power and exhibit higher thermal density than lower-density modules, which tend to remain cooler. Furthermore, modules with heavy dynamic workload activity heat up due to increased switching power and thus age more rapidly, whereas lightly loaded modules stay cooler but experience stronger BTI aging due to prolonged idle bias intervals. Figures \ref{fig:top_module2} and \ref{fig:top_module} illustrate two communicating top-level modules, in two scenarios, A and B, respectively.

We conducted aging simulations using TMI libraries and Cadence Spectre® to evaluate aging effects. The simulation flow consists of three phases: (i) transient simulation in fresh mode, (ii) a stress phase in which aging parameters and operating conditions are applied to extract aged transistor models, and (iii) transient simulation of the aged models to quantify delay degradation. In scenario A, the path between two top modules includes combinational logic, with an initial launch path latency of 1.4 ns comprising clock buffers, flip-flop, and combinational logic, and a corresponding capture path latency of 450 ps. In scenario B, the initial clock insertion delay to the modules is 900 ps. In both scenarios, setup and hold timing constraints are initially closed with zero slack in the fresh design.



\vspace{-23pt}

Table~\ref{tab:exp_setup} summarizes the baseline experimental environment, while Tables~\ref{tab:setup_cases} and~\ref{tab:hold_cases} list the eight evaluated temperature and duty-cycle test cases.

\begin{table}[ht]
\centering
\caption{Summary of the experimental setup used for aging and timing analysis.}
\begin{tabular}{|l|l|}
\hline
\textbf{Parameter} & \textbf{Configuration} \\ \hline

Technology Node & 16\,nm FinFET \\ \hline
Operating Frequency & 1\,GHz \\ \hline
PVT Corner & SS (Slow–Slow) \\ \hline
Cell Types & ULVT clock buffers, SVT standard cells \\ \hline
Duty Cycles Evaluated & 10\% and 50\% \\ \hline
Temperature Conditions &
Moderate: 90\,°C / 105\,°C; \\
& High: 110\,°C / 125\,°C \\ \hline

\end{tabular}
\label{tab:exp_setup}
\end{table}

To evaluate how workload imbalance and temperature gradients jointly influence asymmetric aging, we define four representative operating test cases in both scenarios A and B, varying both power intensity and duty-cycle activity. These test cases capture conditions where a module becomes either thermally stressed or bias-stressed, leading to divergent aging rates. Each one is examined under both setup and hold timing analysis by instantiating it with the corresponding temperature and duty-cycle pairs: (125°C, 110°C) and (105°C, 90°C) combined with 50\% or 10\% activity. The resulting eight test cases for both scenarios, summarized in Tables~\ref{tab:setup_cases}–\ref{tab:hold_cases}, form the basis of our timing evaluation under asymmetric aging conditions.

\vspace{-14pt}
\begin{table}[ht]
\centering
\caption{timing-analysis for scenario A with setup timing slack after aging effect.}
\vspace{-5pt}
\label{tab:setup_cases}
\begin{tabular}{|c|c|c|c|}
\hline
\textbf{Case} &
\textbf{Launch (T, DC)} &
\textbf{Capture (T, DC)} &
\textbf{ Slack} \\ \hline

1 & 125$^\circ$C, 50\% & 110$^\circ$C, 50\% & -41.5ps - \textit{ Setup Violation} \\ \hline
2 & 125$^\circ$C, 50\% & 110$^\circ$C, 10\% & -40.2ps - \textit{Setup Violation} \\ \hline
3 & 105$^\circ$C, 50\% & \;\,90$^\circ$C, 50\% & -32.5ps - \textit{Setup Violation} \\ \hline
4 & 105$^\circ$C, 50\% & \;\,90$^\circ$C, 10\% & -31ps - \textit{Setup Violation} \\ \hline

\end{tabular}
\end{table}

\vspace{-20pt}
\begin{table}[ht]
\centering
\caption{timing-analysis for scenario B with hold timing slack after aging effect.}
\vspace{-5pt}
\label{tab:hold_cases}
\begin{tabular}{|c|c|c|c|}
\hline
\textbf{Case} &
\textbf{Capture (T, DC)} &
\textbf{Launch (T, DC)} &
\textbf{Slack} \\ \hline

5 & 125$^\circ$C, 50\% & 110$^\circ$C, 50\% & -4ps - \textit{Hold Violation} \\ \hline
6 & 125$^\circ$C, 50\% & 110$^\circ$C, 10\% & -2.5ps - \textit{Hold Violation} \\ \hline
7 & 105$^\circ$C, 50\% & \;\,90$^\circ$C, 50\% & -2.1ps - \textit{Hold Violation} \\ \hline
8 & 105$^\circ$C, 50\% & \;\,90$^\circ$C, 10\% & -1.7ps - \textit{Hold Violation} \\ \hline

\end{tabular}
\end{table}
\vspace{-11pt}
All timing analysis in scenario A show significant setup timing slack loss (up to --41.5\,ps), with setup violations occurring whenever the launch path is hotter and thus ages faster. Scenario B,  also exhibit consistent degradation, with hold timing slack reductions of --1.7\,ps to --4\,ps when the capture path ages faster. These results confirm that asymmetric aging driven by temperature and duty-cycle imbalance leads to both setup and hold failures.

\section{Conclusion}
\label{sec:four}
In conclusion, this study demonstrates that asymmetric transistor aging in advanced technology nodes can be substantially amplified by realistic on-chip thermal gradients. Typical temperature gradients of 10-15$^\circ$C produce divergent degradation rates across timing paths, resulting in setup and hold violations that are not captured by conventional uniform-aging assumptions. These findings highlight the need for timing signoff methodologies that explicitly account for both thermal effects and device aging. As design margins continue to shrink in deeply scaled technologies, such enhanced analysis becomes essential for ensuring long-term correctness and reliable system operation.

\bibliographystyle{IEEEtran}
\bibliography{article}
\end{document}